\documentclass[conference]{IEEEtran}
\IEEEoverridecommandlockouts
\usepackage[hyphens]{url}
\usepackage{hyperref}
\usepackage[utf8]{inputenc}
\usepackage{tikz}
\usetikzlibrary{trees,arrows}
\usepackage{url}
\usepackage[hyphenbreaks]{breakurl}
\usepackage{cite}
\usepackage{amsmath,amssymb,amsfonts}
\usepackage{algorithmic}
\usepackage{graphicx}
\usepackage{textcomp}
\usepackage{xcolor}
\usepackage{times}
\usepackage{booktabs}
\usepackage{array}

\def\BibTeX{{\rm B\kern-.05em{\sc i\kern-.025em b}\kern-.08em
    T\kern-.1667em\lower.7ex\hbox{E}\kern-.125emX}}


\usepackage{xpatch}
\xpatchcmd\IEEEkeywords{---}{-}{}{}

\makeatletter
\renewcommand{\fnum@figure}{Figure~\thefigure}
\makeatother

\title{\bfseries\Large Fingerprinting and Tracing Shadows: The Development and Impact of Browser Fingerprinting on Digital Privacy}

\author{\IEEEauthorblockN{Alexander Lawall}
\IEEEauthorblockA{%\textit{dept. name of organization (of Aff.)} \\
\textit{IU International University of Applied Science}\\
Erfurt, Thüringen, Germany \\
alexander.lawall@iu.org}
}

\begin{document}

\maketitle

\begin{abstract}
Browser fingerprinting is a growing technique for identifying and tracking users online without traditional methods like cookies. This paper gives an overview by examining the various fingerprinting techniques and analyzes the entropy and uniqueness of the collected data. The analysis highlights that browser fingerprinting poses a complex challenge from both technical and privacy perspectives, as users often have no control over the collection and use of their data. In addition, it raises significant privacy concerns as users are often tracked without their knowledge or consent.
\end{abstract}

% A list of IEEE Computer Society appoved keywords can be obtained at
% http://www.computer.org/mc/keywords/keywords.htm

\section{Introduction} \label{intro}
%\subsection{Background and Motivation}
In the increasingly digitized world, the issues of online privacy and data security are becoming more complex. Particularly in tracking — monitoring users and their devices across different web servers — browser fingerprinting has emerged as an effective technique for creating detailed user profiles. Unlike the storage of information via cookies, which requires explicit user consent as mandated by the European General Data Protection Regulations (GDPR) guidelines, fingerprinting does not require such consent. A browser fingerprint can be generated in the background without any obvious signs to the end user, leaving them unaware of whether and to what extent they are being tracked.

It is possible to manipulate a device locally to alter its fingerprint. This is often not feasible for all users, unlike deleting cookies. This invisible threat is not apparent to the general public and raises significant privacy concerns, as individuals can be tracked unnoticed. These profiles can contain private information, depending on the server operators, including age group, ethnic origin, social circles, and interests of the affected person.

Browser fingerprinting poses a threat to the privacy of the general public. Contrary to being a threat, it is an opportunity to provide valuable information to enhance the authentication mechanisms. Both perspectives are explored throughout this paper. The focus will be on the various techniques of fingerprinting %, along with an analysis of the entropy and uniqueness of the collected data,
to understand how accurate and detailed user profiles can be created. The main research questions that this paper seeks to answer are: 

%\subsection{Research Question and Objectives}
% How reliable is the identification of individual users?
% What impact does browser fingerprinting have on online tracking of users and their privacy?
% What steps have been taken by external entities to protect the public from fingerprinting?
\begin{itemize}
    \item[RQ1] \textit{``What methods are used in browser fingerprinting and what user data are collected in the process?''}
    \item[RQ2] \textit{``How has the development of browser fingerprinting as a user identification method influenced user privacy and data protection in the digital space?''}
\end{itemize}

The paper is structured as follows: Section \ref{intro} introduces browser fingerprinting and its privacy implications. In Section \ref{theory}, the theoretical background explains how fingerprinting works and its legal challenges. Section \ref{analysis} outlines techniques like HTTP Headers, Canvas, and WebGL Fingerprinting. Section \ref{discussion} examines the impact of fingerprinting on privacy and the regulatory landscape. Section \ref{concl} concludes with a summary of the findings, emphasizing the need for stronger privacy measures and further research on countermeasures.
%\subsection{Contribution and Significance}

\section{Theoretical Background} \label{theory}
\subsection{Fingerprinting}
Browser fingerprinting refers to collecting characteristic information that the browser directly or indirectly reveals about itself. Often used to track users, this technology has also found applications in IT security, such as fraud detection. Unlike tracking methods like cookies, browser fingerprinting does not require storing data on the user's computer, allowing the process to occur secretly and without consent \cite[p. 1]{Mowery2012PixelP}. Consequently, creating a new identity, similar to deleting cookies, is not easily achievable, and GDPR privacy laws often provide little protection. Unlike cookie tracking, browser fingerprinting is not explicitly mentioned in the GDPR. It should fall under the collection of identifiable information but website operators frequently claim ``legitimate interest", enabling such data collection without the user's consent \cite{GDPR2018}.

Active transmission of data is not required for browser fingerprinting, as loading a webpage can transmit various pieces of information, such as the user's preferred language, within the HTTP headers. This passive data collection provides only a limited amount of information, so it is often supplemented with active data collection methods. An active approach typically employs JavaScript to interface with the browser and gather information, such as screen resolution, installed add-ons, and graphics card data, merging them into a unique fingerprint \cite[pp. 1, 3]{Zhang2022}.

Similar to human fingerprints, browser fingerprinting relies on the uniqueness of browser characteristics, which typically do not change significantly with regular use. This allows for accurate user identification over extended periods \cite[p. 2]{Zhang2022}. However, not all collected data points are equally unique or stable, necessitating careful selection of information to achieve accurate results. The fingerprinting algorithm combines both passively and actively collected data into a unique string. Depending on the operator's goals, adjustments can be made; for instance, using cookies, the fingerprint might be less stable but more unique, while tracking users without cookies requires high stability \cite[pp. 1-5]{Eckersley2010}. Eckersley's study showed that participant browsers already had high entropy, indicating many unique characteristics sufficient for accurate fingerprinting, though not stable enough for long-term accuracy. In recent years, potential entropy has increased with new techniques like HTML Canvas, WebGL-based hardware fingerprints, audio API fingerprints, plug-in-based fingerprints, and methods utilizing mouse movements or differences in HTML parsing between browsers, making cross-browser user identification possible \cite[pp. 4-5]{Zhang2022}.

\subsection{Concerns for Digital Privacy}
Historically, the greatest threat to online tracking was posed by cookies, along with other technologies like Flash cookies, which have lost significance in recent years. Changes by browser manufacturers, such as Mozilla, which rendered many exploited technologies, so-called ``super-cookies", ineffective \cite{Firefox2021}, and additional browsers planning to block or eliminate third-party cookies in the coming years \cite{Woollacott2021}, have shifted the landscape. Following the GDPR, the use of non-essential cookies has been further restricted and standardized for the first time, defining how users share their data through cookies \cite{CookieDirectiveGDPR}. In contrast, browser fingerprinting occurs in the background and leaves no stored information on the user's computer. Thus, the use of fingerprints not only circumvents previous issues related to local storage, such as privacy laws and technical limitations but also persists even when local data is deleted or when incognito mode is used.

A 2021 study of the Alexa Top 100,000 websites found that nearly 10\% of the sites used scripts to generate fingerprints \cite[pp. 11-12]{Iqbal2021}. Comparing this to a similar 2014 study, which recorded 5.5\% of the top 100,000 sites using canvas fingerprinting scripts, reveals an almost doubling of usage over seven years \cite{Acer2014}. This suggests a shift towards online tracking using this technology, which is much harder to detect and prevent compared to cookies. The creation of a fingerprint is imperceptible to the user, with no simple way to effectively change or delete their fingerprint. Cookie banners give a false sense of security while tracking continues in the background without consent.

Thus, browser fingerprinting poses an active threat to privacy, as users often have no control over the collection and use of their data. This stands in opposition to many current data protection principles, such as the GDPR.

\section{Methods of Browser Fingerprinting}\label{analysis} % Results and Technics
In the context of browser fingerprinting techniques, the methods of data collection are varied and comprehensive. Therefore, specific properties and criteria are used to select techniques. The following sections will encompass the explanation of the techniques in terms of their functionality and their applications will be discussed to provide a detailed understanding of their use. An evaluation based on the advantages and disadvantages of each technique is also included to weigh their effectiveness and potential risks. Given the ever-increasing number of techniques, only the most commonly used, established, or novel methods will be presented here.

\subsection{HTTP Header Attributes} \label{sec:http-header-attributes}
\subsubsection{Definition and Basics}
The HTTP request header is a part of every HTTP request exchanged between a client (web browser) and a server, transmitting various functional and compatibility-related information \cite{ReqHead24}. Although individual attributes are not unique, they can be combined to distinguish a client. This explanation is based on HTTP version 1.1, with HTTP/2 maintaining most attributes within a modified header frame \cite{Http224}.
\subsubsection{Analysis}
HTTP request headers include attributes that differ by browser and version. Effective fingerprinting requires selecting attributes that remain consistent over time. Reliable fields include User-Agent, Accept, Content-Encoding, and Content-Language, which provide valuable identification information \cite[p. 5]{Eckersley2010}\cite[p. 880]{Laperdrix2016}. The User-Agent, despite lacking standardization, offers high uniqueness due to its detailed browser and OS information \cite{UserAg24}.

\subsubsection{Advantages}
The main advantage of using HTTP headers is their passive information collection, which occurs automatically with each request. This method is efficient, unobtrusive, and compatible with most web servers, processing data on the server side without a noticeable impact on the client.
\subsubsection{Disadvantages}
HTTP header information is limited, as most attributes provide minimal details. The User-Agent, while informative, can be easily altered by browser extensions, reducing its reliability (i.e. User-Agent Switcher for Chrome). Furthermore, using such technologies without consent can violate GDPR regulations, necessitating legal review before implementation \cite{Wolford2024}.

\subsection{Enumeration of Browser Plugins} \label{sec:browser-plugins}
\subsubsection{Definition and Basics}
Browser plugins, whether pre-installed or user-added, have been a method for recognizing systems, along with font detection. Most browser features are indirectly modified, except for extensions. The demand for accurate enumeration of these extensions is high \cite[pp. 878-880]{Laperdrix2016}.
\subsubsection{Analysis}
Many information-rich plugins, like Flash, have disappeared over the years. Since 2016, most browsers, including Firefox, no longer support the Netscape Plugin Application Programming Interface (NPAPI) plugin interface, leading to the navigator.plugins object in modern browsers showing only standard plugins like PDF viewers \cite{Navi24}. This limitation reduces the impact of plugins on fingerprinting but still allows differentiation between systems and browsers. The direct detection of user-installed add-ons is not possible, limiting the data's significance \cite[pp. 886-887]{Laperdrix2016}. However, new methods to enumerate extensions have emerged. Chromium-based browsers can access extension settings via a local URL. A GitHub project exploits this to check for over 1,000 extensions by requesting internal resources and checking the status codes \cite{Extdetect24}. Additionally, ad blockers' behavior in removing unwanted content can be detected by creating elements they typically block and checking for changes, revealing active blocklists \cite{FiBlocker24}. Another method involves reading the status of handler protocols to identify installed programs like Skype and Zoom.

\subsubsection{Advantages}
User-installed extensions offer high uniqueness and stability due to the number of extensions.
\subsubsection{Disadvantages}
Insights into users' privacy, including sensitive information like health conditions, religion, and political views, can be inferred \cite[pp. 11-12]{Karami2020}. The fingerprinting process relies on limited methods, making it prone to errors, and requires continuous updates to maintain reliability.

\subsection{Canvas Fingerprinting} \label{sec:canvas-fingerprinting}
\subsubsection{Definition and Basics}
Canvas fingerprinting involves generating a digital fingerprint using the Canvas element introduced in HTML5. It utilizes the Canvas API to draw a hidden 2D graphic in the background. Variations in how different browsers and devices handle this image due to differences in hardware acceleration, installed fonts, and graphic libraries result in a highly stable and unique fingerprint \cite[pp. 1-3]{Mowery2012PixelP}.
\subsubsection{Analysis}
A script embedded in a webpage adds an invisible Canvas element that draws a predetermined 2D graphic in the background. Text can also be drawn using the Canvas context, employing various fonts and sizes. WebFonts enable dynamic loading of fonts from the internet, allowing specific fonts to be chosen to test for uniqueness in font rendering. The resulting image data can be extracted using functions like \textit{getImageData} and \textit{toDataURL}, which can then be hashed to form a fingerprint, typically using a hashing algorithm. This hash is sent via a web request to a server for processing and storage. Besides storing the fingerprint for later identification, another application method involves comparing the fingerprint with an extensive database of known fingerprints and corresponding system configurations, enabling reliable system profiling \cite[pp. 2-4]{Mowery2012PixelP}.

\subsubsection{Advantages}
Mowery and Shacham demonstrated that implementing Canvas fingerprinting is straightforward, requiring minimal lines of client-side code. It leverages basic JavaScript functions and can be deployed across all major web applications. The fingerprinting process is discrete for users and challenging to block because Canvas operations are common in web applications, making it difficult to distinguish normal operations from fingerprinting scripts. The simplicity of fingerprint creation enables high speed, stability, uniqueness, and entropy, making it particularly valuable for real-time tracking applications \cite[pp. 1-5]{Mowery2012PixelP}.
\subsubsection{Disadvantages}
Changes in browser environments, such as updates or graphic settings, can affect the stability of the fingerprint. Variability in hardware and software configurations can lead to inconsistencies. As an active technique, executing code on the client side is necessary, posing risks of detection and potential blockage by blocklists targeting known fingerprinting scripts \cite[pp. 3-7]{Mowery2012PixelP}. While imperceptible to users, the limited interfaces to retrieve generated Canvas data can be monitored and manipulated by extensions. Add-ons like CanvasBlocker allow users to prevent data retrieval or manipulate Canvas data, continuously generating new fingerprints to prevent identification \cite{CanvasBlocker24}. Finally, while implementing Canvas fingerprinting is ``relatively simple", analyzing and interpreting the data can be complex and may require expertise in the field \cite[pp. 6-8]{Mowery2012PixelP}.

\subsection{WebGL Fingerprinting} \label{sec:webgl-fingerprinting}
\subsubsection{Definition and Basics}
WebGL fingerprinting is a technique utilizing the WebGL JavaScript API, based on OpenGL ES 2.0, allowing web applications to render both 2D and 3D graphics with high performance by directly accessing the GPU \cite{WebGL24}. Unlike Canvas fingerprinting, which focuses on 2D graphics and identifies software differences mainly through fonts and graphic libraries, WebGL fingerprinting provides deeper and more precise detection capabilities. It captures unique hardware information, particularly details about the graphics processor, distinguishing it significantly from Canvas fingerprinting and broadening its application for tracking purposes \cite[p. 4]{Mowery2012PixelP}.
\subsubsection{Analysis}
WebGL fingerprinting uses a Canvas element to access the API. Similar to Canvas fingerprinting, it creates an invisible element performing 3D operations in the background to collect data without user interaction. A straightforward application involves accessing specific variables, such as \textit{UNMASKED\_VENDOR\_WEBGL} and \textit{UNMASKED\_RENDERER\_WEBGL}, using the \textit{getParameter} function in the WebGL context. These variables provide information about the graphics hardware manufacturer (Vendor) and model (Renderer). For example, a Vendor entry like ``Intel" indicates an integrated graphics unit, while ``Nvidia" combined with ``GeForce GTX 970" as Renderer indicates a dedicated graphics card. These details can reveal insights into the system being used \cite[p. 17]{Stephenson2023}. Privacy concerns have led browsers like Apple's WebKit to provide generic information instead of specific data to protect user privacy. Since 2020, WebKit has masked Vendor and Renderer information, as well as shading language details \cite{WebKit24}. Firefox similarly groups graphics processor models into categories instead of displaying specific models. In practice, this means that a Nvidia card from the 900 series onward, for example, is reported as ``GeForce GTX 980". In summary, research investigating hardware fingerprinting using HTML5 demonstrated the capability to identify devices based on GPU performance. It utilizes the graphics processor's clock frequency and clock skew to render complex 3D graphics, measuring GPU performance based on the number of frames rendered within a period, providing insights into the GPU's frequency and core count \cite[pp. 3-4]{Nakibly2015}.

\subsubsection{Advantages}
As demonstrated by Cao et al., WebGL can offer high uniqueness and stability \cite{Cao2017CrossBrowserFV}. Its direct interface with the system ensures consistency across browsers, making it challenging for users to evade identification through simple browser changes or reinstalls. Despite changes to enhance WebGL's resistance to fingerprinting, it reliably identifies users. The successor to WebGL, WebGPU, is currently in development, promising even more privacy risks due to its closer hardware access, allowing for classifications with up to 98\% accuracy in 150 milliseconds, a reduction from the 8 seconds WebGL took \cite{Mantel2022}.
\subsubsection{Disadvantages}
The complexity of WebGL fingerprinting is significantly higher compared to previous techniques, necessitating careful consideration whether a simpler Canvas approach combined with other methods might be accurate enough for specific use cases. Intensive tasks in a 3D environment can also strain the target system, leading to longer fingerprint creation times \cite[p. 4]{Mowery2012PixelP}. Implementing WebGL requires caution, as shown by the cases of Laperdrix et al. and Cao et al., and opting for a ready-made solution might be advisable. Moreover, WebGL shares Canvas's vulnerability to blocked or misread data if detection methods rely on differences in rendered graphics. Even novel methods like DrawnApart can be mitigated through countermeasures, such as limiting to a single EU \cite[p. 12]{Laor2022DRAWNAPARTAD}. WebGL may also not be available or disabled on some devices, necessitating consideration of alternatives, such as using the 2D Canvas.

\subsection{Audio Fingerprinting} \label{sec:audio-fingerprinting}
\subsubsection{Definition and Basics}
The Web Audio API is a JavaScript interface for processing and synthesizing audio signals in the web browsers, part of the HTML5 standard. It can identify systems through manufacturing differences in audio hardware. Methods analyze signal processing characteristics, hardware differences, and system responses to specific audio signals for fingerprinting \cite[pp. 1107-1109]{Queiroz2019AWB}.
\subsubsection{Analysis}
Audio fingerprinting involves various acoustic measurements to create a unique device fingerprint. It requires an AudioContext linking an AudioBuffer, Oscillator, and Compressor. The AudioBuffer represents a small audio segment, while the Oscillator generates a waveform at a defined frequency. The Compressor manipulates the audio signal. The unique waveform generated and manipulated reflects system characteristics, allowing a unique fingerprint to be created using a hash function on the final waveform. This method, known as ``Dynamic Compressor (DC)", is highly stable, producing the same fingerprint for the user each time using a reliable hash function \cite[pp. 1109-1111]{Queiroz2019AWB}.

Another method is the ``Fast Fourier Transform" (FFT), converting audio signals from the time domain to the frequency domain. It measures hardware implementation differences to identify characteristics. FFT is less stable than DC, often requiring multiple attempts for consistent results. DC and FFT are often used together for more reliable outcomes \cite[pp. 1111-1114]{Queiroz2019AWB}.
Researchers compared the techniques, including custom-designed ones, alongside DC and FFT. These included creating ``Custom Signals", ``Merged Signals", and analyzing generated AM and FM waves. All techniques showed good stability, averaging two to four attempts for fingerprint matching \cite[pp. 3-5]{Chalise2021}.

\subsubsection{Advantages}
The generated fingerprints are highly stable and can differentiate systems based on their properties. Queiroz and Feitosa showed that mobile devices using Firefox could be consistently recognized and grouped by their stable fingerprints \cite[p. 1119]{Queiroz2019AWB}. Techniques like DC are simple to implement and offer high stability. Other promising techniques, especially when used together, could enhance potential but are more challenging to implement \cite[pp. 1-3]{Chalise2021}.
\subsubsection{Disadvantages}
While audio fingerprinting offers high stability, it lacks uniqueness and accuracy on its own and should be used with other fingerprinting techniques \cite[p. 1119]{Queiroz2019AWB}. Additionally, the Web Audio API can be disabled on devices or manipulated by add-ons like ``Canvas Blocker", which also blocks and manipulates Canvas and WebGL.

\subsection{Font Fingerprinting} \label{sec:font-fingerprinting}
\subsubsection{Definition and Basics}
Font fingerprinting is a browser fingerprinting technique that identifies devices by recognizing installed fonts. This method creates unique digital fingerprints by combining fonts with other data points, which can be used for tracking and identification purposes \cite[p. 314]{Gomez2018}.
\subsubsection{Analysis}
After the end of Adobe Flash, a new method for font recognition was needed. JavaScript uses a fallback mechanism to recognize fonts by comparing the dimensions of texts in specific fonts with expected values. Invisible \textit{div} elements and the canvas element are used to identify installed fonts \cite[p. 311]{Gomez2018}\cite[p. 12]{Englehardt2016}. The experimental Local Font Access API requires user consent and is therefore not suitable for fingerprinting \cite{FontAPI24}.

\subsubsection{Advantages}
Font recognition offers high entropy and stability since fonts are rarely changed. This allows the identification of the operating system and installed software packages like Office or Photoshop \cite[p. 7]{Zhang2022}.
\subsubsection{Disadvantages}
Without Flash, font recognition is done through ``brute-force" methods, reducing accuracy if unknown fonts are installed. Similar fonts can lead to false positives. Extensions and adjustments, such as those in Apple's WebKit, can manipulate or restrict recognition \cite[p. 10]{Cao2017CrossBrowserFV}\cite[p. 311]{Gomez2018}.

\subsection{Screen Fingerprinting} \label{sec:screen-fingerprinting}
\subsubsection{Definition and Basics}
Screen fingerprinting identifies a device by analyzing various screen-related characteristics, including screen resolution, pixel depth, color depth, and browser window size. This method leverages the uniqueness of screen configurations and browser modifications, which can create rare resolution combinations \cite[p. 20]{Laperdrix2020}.
\subsubsection{Analysis}
JavaScript provides attributes for screen and browser window characteristics through the \textit{window.screen object}, offering details like color depth (\textit{colorDepth}), screen orientation (\textit{screenOrientation}), and screen dimensions (\textit{screenHeight}, \textit{screenWidth}). Values, such as \textit{window.innerWidth} and \textit{window.innerHeight}, determine the browser window's inner area, which can be altered by toolbars or bookmark bars \cite[p. 3]{Cao2017CrossBrowserFV}.

\subsubsection{Advantages}
Screen and window resolution information typically have high entropy, making them useful for stabilizing fingerprints when combined with other techniques. This method is particularly effective for distinguishing between desktop, tablet, and mobile devices, as these have distinct resolutions and aspect ratios compared to standardized desktop screens \cite[p. 277]{Queiroz2019AWB}.
\subsubsection{Disadvantages}
Since values are derived from browser attributes rather than hardware tests, they can be limited or altered by extensions or privacy settings. Browsers like TOR set the window to a fixed size of 1000x1000 pixels, reducing uniqueness, and browsers like Firefox always report a color depth of 24. Additionally, users with multiple monitors or those using zoom functions can affect the accuracy of screen fingerprinting, as there is no reliable way to determine the zoom factor directly, which reduces entropy \cite[p. 10]{Cao2017CrossBrowserFV}.

\subsection{WebRTC Fingerprinting} \label{sec:webrtc-fingerprinting}
\subsubsection{Definition and Basics}
WebRTC is a standard and accessible JavaScript interface available in most browsers. It facilitates real-time communication over stateless HTTP by establishing direct connections between participants, allowing the extraction of local network adapter information. This can reveal private and public IP addresses, which can be used for fingerprinting or identifying users behind proxies or VPNs \cite[p. 12]{Englehardt2016}. It also provides information about connected devices, such as microphones, webcams, and speakers.
\subsubsection{Analysis}
Unlike other browser mechanisms like camera or microphone access, establishing a WebRTC connection requires no permissions or user notifications. After connecting to the target computer via a Session Traversal Utilities for NAT (STUN) server, IP addresses can be read from the RTCPeerConnection object as iceCandidates \cite[p. 667]{Reiter2017}. This data can be used for fingerprinting, and WebRTC can further enumerate the local network to build a unique profile of the target's environment. It can also read all local adapter addresses, including those for VPNs and virtual machines \cite[p. 667-668]{Reiter2017}.
The DetectRTC project \cite{DetectRTC24} demonstrates WebRTC's capabilities, highlighting information about microphones, webcams, and speakers. While exact device names require permissions, WebRTC can read Media Device IDs, which can contribute to unique fingerprints.

\subsubsection{Advantages}
Extracting private and public IPs provides deep insights, especially for identifying targets behind VPNs or proxies. No other technique can silently reveal addresses behind Network Address Translation (NAT) \cite[p. 273]{Bernardo2016}. The collected data is highly unique; a study with 80 devices found over 97\% uniqueness using only WebRTC \cite[p. 668]{Reiter2017}.
\subsubsection{Disadvantages}
WebRTC might be disabled in the target browser, or extensions might block its usage without user consent. Accessing Media Device IDs requires permission, alerting users to potential background activities, making it unsuitable for stealth operations. Additionally, WebRTC relies on STUN servers, either self-hosted or third-party, adding dependency considerations for its use.

\subsection{CSS Fingerprinting} \label{sec:css-fingerprinting}
\subsubsection{Definition and Basics}
Different to the active fingerprinting techniques using JavaScript, CSS fingerprinting is a passive method. CSS is a stylesheet language primarily used to enhance the presentation of HTML elements. Over time, the CSS specification has expanded to include selectors and filters, enabling limited dynamic selections, which this technique leverages \cite[p. 10]{Bujlow2017}.
\subsubsection{Analysis}
Until 2010, the \textit{:visited} selector could identify if a website had been visited by changing the link color, detectable via JavaScript. After this was patched, researchers explored time-based methods to read user history, but these required JavaScript and were impractical \cite[p. 4]{Olejnik2012}. In 2015, Takei et al. introduced a JavaScript-free method using CSS properties and multiple \textit{@media} queries to fetch URLs based on defined rules. The server could then identify system properties like screen dimensions, resolution, touchscreen presence, installed fonts, browser, and OS from the requesting IP address and URL parameters \cite[p. 3-5]{Takei2015}. A current GitHub project demonstrates this method's practical capabilities \cite{CSSFingerp24}.

\subsubsection{Advantages}
CSS fingerprinting's independence from JavaScript allows it to identify even cautious users who block JavaScript or use extensions like NoScript. This technique can even detect if JavaScript is disabled via noscript tags \cite[p. 2]{Takei2015}. Due to its limited use and lesser-known status, no effective user solutions currently exist to prevent it.
\subsubsection{Disadvantages}
Takei et al.'s method provides limited data, which, without JavaScript, can only be supplemented by techniques like header analysis. Oliver Brotchie notes in his project repository that the method is not currently scalable, as each request requires over 1MB of CSS files to be downloaded. However, he warns that upcoming CSS Values 4 implementation could reduce download sizes significantly, making the method more practical. Additionally, font recognition relies on brute-forcing, which can be noticeable in network traffic.

\subsection{Additional JavaScript Attributes} \label{sec:javascript-attributes}
\subsubsection{Definition and Basics}
Most of the previously discussed techniques actively use JavaScript to extract information from various interfaces. Additional possibilities are briefly mentioned here to provide a more comprehensive picture. Since these techniques share many characteristics with other JavaScript-based methods, listing their pros and cons is omitted.
\subsubsection{Analysis}
The \textit{navigator} object in browsers provides information, such as DoNotTrack status, user agent details, platform, languages, cookies usage, granted and available permissions, and time zone \cite[p. 9]{Gomez2018}. JavaScript implementation varies between browsers and versions, and Mowery et al. demonstrated that these differences are measurable and can indicate the software and hardware used \cite{Mowery2012PixelP}.

Additionally, there are differences in the availability and execution of functions, which offers an alternative way to detect user agents if manipulated by extensions. Another technique that caused concern among Tor users is the use of the \textit{getClientRects} function to obtain precise DOM element data, even with Canvas disabled. These factors can vary based on implementation, font sizes, and screen resolutions, enabling identification in the otherwise anonymous browser \cite{AdvTorFiP24}. This vulnerability has been fixed in Tor but remains exploitable in other browsers \cite{GetClient24}.

\subsubsection{Advantages}
JavaScript-based fingerprinting techniques are highly versatile and widely applicable since JavaScript is essential for web functionality. These methods can collect a broad range of information, such as user agent details, time zones, and system settings, often without requiring user consent or visibility. The stealthy nature of JavaScript fingerprinting allows it to operate in the background, making it difficult for users to detect. Moreover, JavaScript-based attributes work consistently across different browsers, enabling effective cross-browser tracking.

\subsubsection{Disadvantages}
However, JavaScript fingerprinting is limited by browser-specific implementations, which can result in inconsistent data collection. Privacy-focused browsers like Tor or extensions, such as NoScript, actively block or obscure JavaScript-based tracking, reducing its effectiveness. Additionally, users are becoming more aware of privacy risks and increasingly use tools to disable or modify JavaScript functions. Finally, updates to browsers may close vulnerabilities or alter features that JavaScript fingerprinting relies on, decreasing its long-term viability.

\subsection{Advanced Techniques Using Machine Learning} \label{sec:ml-techniques}
\subsubsection{Definition and Basics}
Most active techniques discussed so far use JavaScript to gather hardware and software information. They rely on unique data combinations based on implementation quirks or directly available information. Newer methods often employ ``side-channels", capturing additional data by observing behavioral differences during various operations within the execution environment. Methods like plugin enumeration (cf. Section \ref{sec:browser-plugins}), font fingerprinting (cf. Section \ref{sec:font-fingerprinting}), and CSS fingerprinting (cf. Section \ref{sec:css-fingerprinting}) use this approach in simple forms by testing known combinations to gain indirect information. These side-channel methods can be implemented with minimal effort but can also be used in more sophisticated ways with machine learning to gather otherwise unobtainable information \cite[p. 1]{Wang2021}.
\subsubsection{Analysis}
Wang et al. explored using cache usage, memory consumption, and CPU activity to identify visited websites. Previously, CSS selectors were used to reveal browsing history, posing significant privacy risks and leading to prompt fixes. Side-channel techniques employ various tricks to analyze system behavior more accurately. Complex calculations stress the hardware in the background, and machine learning models categorize the results with expected values from known sites. Tests showed 80-90\% accuracy in identifying websites \cite[p. 3-5]{Wang2021}. Further research is needed, but implementations using WebAssembly \cite{WebAssembly24} and the Performance API \cite{PrecTiming24} are conceivable.

\subsubsection{Advantages}
This method is invisible to the user and provides insightful information not available through conventional means. Currently, there are no methods to protect users from such techniques \cite[pp. 1-3]{Wang2021}.
\subsubsection{Disadvantages}
While previous techniques aimed to identify a user over time, this method could offer dangerous insights into the person's behavior behind the screen. However, the technique is still in its initial stage and remains a theoretical approach not yet tested in the real world. It is unlikely to be reliably used by actors in the near future \cite[p. 6]{Wang2021}.

\section{Discussion} \label{discussion} % \textbar{} Evaluation}
% How reliable is the identification of individual users?
% What impact does browser fingerprinting have on online tracking of users and their privacy?
% What steps have been taken by external entities to protect the public from fingerprinting?

\begin{table*}
    \centering
    \caption{Overview of Fingerprinting Methods}
    \begin{tabular}{|m{3cm}|m{1.4cm}|m{1.4cm}|m{1.4cm}|m{4cm}|m{4cm}|}
        \hline
        \textbf{Fingerprinting Method}      & \textbf{Uniqueness} & \textbf{Stability} & \textbf{Entropy} & \textbf{Impact on User Privacy} & \textbf{Defense Techniques} \\ \hline
        \hline
        \textbf{HTTP Header Attributes}     & Low            & Moderate      & Low         & Moderate impact: limited detail but useful when combined with other methods. & Altering or masking headers (e.g., randomizing User-Agent). \\ \hline
        \textbf{Enumeration of Browser Plugins} & Moderate        & High          & High         & High impact: reveals sensitive data, such as installed plugins. & Disabling plugin enumeration, avoiding unnecessary add-ons. \\ \hline
        \textbf{Canvas Fingerprinting}       & High            & Moderate      & High         & High impact: generates unique fingerprints based on rendering. & CanvasBlocker extension to block or manipulate rendering. \\ \hline
        \textbf{WebGL Fingerprinting}        & High            & High          & High         & High impact: collects detailed hardware data for tracking. & Block or manipulate WebGL outputs. \\ \hline
        \textbf{Audio Fingerprinting}        & Moderate        & High          & Moderate     & High impact: captures unique audio processing details. & Disable Web Audio API, use privacy extensions. \\ \hline
        \textbf{Font Fingerprinting}         & High            & High          & Moderate     & High impact: identifies installed fonts, making it persistent. & Limit font access with privacy-focused browsers (e.g., Tor). \\ \hline
        \textbf{Screen Fingerprinting}       & Moderate        & High          & Low          & Moderate impact: uses screen resolution and window size but less effective on mobile devices. & Fix window size or limit resolution reporting with privacy browsers. \\ \hline
        \textbf{WebRTC Fingerprinting}       & Very High       & High          & Very High    & Very high impact: exposes real IP addresses, even behind VPNs. & Disable WebRTC, use extensions that block data collection. \\ \hline
        \textbf{CSS Fingerprinting}          & Low             & Moderate      & Low          & Low impact: provides limited system and style information. & Limit or disable CSS fingerprinting through extensions or scripts. \\ \hline
        \textbf{JavaScript Attributes}       & Moderate        & High          & Moderate     & Moderate impact: uses various browser features for tracking. & Disable unnecessary JavaScript functions or use privacy extensions. \\ \hline
        \textbf{Advanced Machine Learning Fingerprinting} & Very High       & Very High     & Very High    & Very high impact: uses side-channel data (e.g., CPU/cache) for tracking. & Limit access to Performance API and WebAssembly, emerging defenses needed. \\ \hline
        \end{tabular}
%    \vspace{0.2cm}
    \label{tab:overview}
\end{table*}

% Unnoticed Data Collection
Browser fingerprinting can be used positively for security, as shown by technologies like BrFast and private, passive user recognition methods. However, there's a risk of misuse, especially in advertising. Personalized ads significantly impact Generation Z, who discover products primarily through social media. The advertising industry, driven by creating accurate user profiles, heavily invests in digital advertising, with data-driven ads accounting for 60-70\% of digital ad revenue in Germany. Traditionally, data collection relied on cookies, but users developed ways to avoid tracking, such as deleting cookies or using incognito mode. Unlike cookies, browser fingerprints are collected in the background and are not easily altered. GDPR regulations mandate user consent for data collection, but enforcement is inconsistent, and compliance with fingerprinting guidelines remains unclear, even with new laws like Germany's TTDSG \cite{DSGVOFP24}.

% Affected Demographics
Online tracking is ubiquitous, affecting nearly all user groups. A 2016 study of the top 1 million websites revealed extensive tracking, with services like Google and Facebook present on over 10\% of sites. Post-GDPR, fingerprinting scripts increased to 68.8\% of the top 10,000 sites. A study with 234 participants found that demographics like age, gender, education, IT background, and privacy awareness influenced trackability, with men and those with higher education being less trackable. Despite understanding fingerprinting, many participants believed they could protect themselves from it. The AmIUnique study, with over 100,000 fingerprints, indicated a bias towards more privacy-aware internet users. Current research from Friedrich-Alexander-University shows that most study participants are male and well-educated, suggesting that while almost everyone is affected by browser fingerprinting, only a small, informed group actively researches and understands it \cite{Pugliese2020}.

% discussion on the greater picture of browser fingerprinting that somehow puts the presented methods in relation to each other
Browser fingerprinting, as explored through various methods in this paper (cf. Table \ref{tab:overview}), represents a comprehensive and evolving threat to digital privacy. Each fingerprinting technique, from HTTP Header Attributes to more sophisticated approaches like Canvas and WebGL Fingerprinting, offers unique data points, but their power lies in their combinatorial use. While individual methods may not be highly unique or stable, their integration enables more persistent and accurate user identification across devices and browsers. Techniques like WebRTC and Font Fingerprinting complement traditional methods by exposing additional layers of system and network data. Furthermore, the advancement of machine learning-based fingerprinting is pushing the boundaries of tracking, allowing for the analysis of side-channel behaviors, such as CPU or memory usage. This convergence of methods creates a powerful, multi-dimensional profiling system that is increasingly resistant to countermeasures, challenging both privacy frameworks and user efforts to remain anonymous online. Therefore, the future of browser fingerprinting lies in this synergistic exploitation of both passive and active methods, making it a critical issue in the broader context of digital surveillance and privacy regulation.

\section{Conclusion} \label{concl}
\subsection{Summary of the Research Outcome}

This contribution has examined browser fingerprinting, a growing technique in online tracking. It has demonstrated that browser fingerprinting is a sophisticated method for identifying and tracking users online without traditional methods like cookies.

The analysis highlighted that browser fingerprinting poses a complex challenge from both technical and privacy perspectives. While it provides companies and advertisers with detailed insights into user behavior for targeted advertising, it raises significant privacy concerns as users are often tracked without their knowledge or consent. Despite stricter privacy laws like the GDPR in the EU, browser fingerprinting remains a grey area. Anti-fingerprinting techniques are limited and continually evolving to keep up with new tracking methods.

In conclusion, browser fingerprinting plays and will continue to play a significant role in the digital landscape. Both users and regulatory bodies must increase awareness of browser fingerprinting practices and their implications. %Future research should focus on developing more effective privacy techniques to balance commercial interests and user privacy rights.

\subsection{Implications for Practice}
% What additional protection measures are available? Which of these can be implemented by average individuals without extensive IT knowledge?
\textbf{Consent and Cookies:} Always accept only the necessary cookies in cookie banners and regularly delete cookies to hinder tracking and fingerprinting. This is particularly important for news sites, which often misuse collected data without user consent.

\textbf{Blending in with the Masses:} Reducing APIs and data sources for fingerprinting can ironically make users more identifiable \cite{Al-Fannah2020}. Thus, widely adopted browsers and protection mechanisms should be used to stay less conspicuous.

\textbf{Browser Choice:} Choose browsers with robust privacy protections. On iOS, Safari is recommended due to its advanced tracking protection and large user base \cite{Kollnig2022}. For Android, the Mull browser is highly rated for fingerprinting protection, while Brave is a good, widely-used alternative. On desktops, Brave, Librewolf, and Mullvad browsers are recommended for their privacy features and user bases \cite{Lin2023}.

\textbf{Browser Extensions:} Limit the use of browser extensions, as they can become sources of unique information. While some extensions block known trackers or modify API outputs, these protections are often already built into recommended browsers like Brave and Librewolf \cite{Karami2020}\cite{Al-Fannah2020}. 

\subsection{Future Research}
Future research in browser fingerprinting should focus on several key areas. First, countermeasures and defense mechanisms need to be explored further, especially in mitigating the newer techniques that leverage machine learning and side-channel attacks. These advanced methods can bypass traditional privacy safeguards, such as disabling JavaScript or using incognito modes, making the development of more robust anti-fingerprinting technologies imperative. Additionally, research should explore the ethics and regulatory frameworks surrounding fingerprinting, examining how existing privacy and data protection laws like GDPR can be adapted to better address fingerprinting practices. Another promising direction is improving cross-device tracking prevention by understanding how fingerprinting works across different platforms and hardware. Lastly, investigating user awareness and educational tools on fingerprint privacy risks will help empower the general public to protect their digital identities more effectively. Thus, future research should focus on developing more effective privacy techniques to balance commercial interests and user privacy rights.

% ======== References =========
\bibliographystyle{IEEEtran} 
\bibliography{references} 

% Generated by IEEEtran.bst, version: 1.14 (2015/08/26)
\begin{thebibliography}{10}
\providecommand{\url}[1]{#1}
\csname url@samestyle\endcsname
\providecommand{\newblock}{\relax}
\providecommand{\bibinfo}[2]{#2}
\providecommand{\BIBentrySTDinterwordspacing}{\spaceskip=0pt\relax}
\providecommand{\BIBentryALTinterwordstretchfactor}{4}
\providecommand{\BIBentryALTinterwordspacing}{\spaceskip=\fontdimen2\font plus
\BIBentryALTinterwordstretchfactor\fontdimen3\font minus
  \fontdimen4\font\relax}
\providecommand{\BIBforeignlanguage}[2]{{%
\expandafter\ifx\csname l@#1\endcsname\relax
\typeout{** WARNING: IEEEtran.bst: No hyphenation pattern has been}%
\typeout{** loaded for the language `#1'. Using the pattern for}%
\typeout{** the default language instead.}%
\else
\language=\csname l@#1\endcsname
\fi
#2}}
\providecommand{\BIBdecl}{\relax}
\BIBdecl

\bibitem{Mowery2012PixelP}
K.~Mowery and H.~Shacham, ``{Pixel perfect: Fingerprinting canvas in HTML5},''
  \emph{Proceedings of W2SP}, vol. 2012, 2012.

\bibitem{GDPR2018}
\BIBentryALTinterwordspacing
K.~Szymielewicz and B.~Budington. (2018) {The GDPR and Browser Fingerprinting:
  How It Changes the Game for the Sneakiest Web Trackers}. Accessed:
  2024-09-27. [Online]. Available:
  \url{https://www.eff.org/de/deeplinks/2018/06/gdpr-and-browser-fingerprinting-how-it-changes-game-sneakiest-web-trackers}
\BIBentrySTDinterwordspacing

\bibitem{Zhang2022}
\BIBentryALTinterwordspacing
D.~Zhang, J.~Zhang, Y.~Bu, B.~Chen, C.~Sun, and T.~Wang, ``{A Survey of Browser
  Fingerprint Research and Application},'' \emph{Wireless Communications and
  Mobile Computing}, vol. 2022, no.~1, p. 3363335, 2022. [Online]. Available:
  \url{https://onlinelibrary.wiley.com/doi/abs/10.1155/2022/3363335}
\BIBentrySTDinterwordspacing

\bibitem{Eckersley2010}
P.~Eckersley, ``How unique is your web browser?'' in \emph{Privacy Enhancing
  Technologies}, M.~J. Atallah and N.~J. Hopper, Eds.\hskip 1em plus 0.5em
  minus 0.4em\relax Berlin, Heidelberg: Springer Berlin Heidelberg, 2010, pp.
  1--18.

\bibitem{Firefox2021}
\BIBentryALTinterwordspacing
S.~Englehardt and A.~Edelstein. (2021) {Firefox 85 Cracks Down on
  Supercookies}. Accessed: 2024-09-27. [Online]. Available:
  \url{https://blog.mozilla.org/security/2021/01/26/supercookie-protections/}
\BIBentrySTDinterwordspacing

\bibitem{Woollacott2021}
\BIBentryALTinterwordspacing
E.~Woollacott. (2021) {Browser fingerprinting more prevalent on the web now
  than ever before}. Accessed: 2024-09-27. [Online]. Available:
  \url{https://portswigger.net/daily-swig/browser-fingerprinting-more-prevalent-on-the-web-now-than-ever-before-research}
\BIBentrySTDinterwordspacing

\bibitem{CookieDirectiveGDPR}
\BIBentryALTinterwordspacing
R.~Koch. (2019) {Cookies, the GDPR, and the ePrivacy Directive}. Accessed:
  2024-09-27. [Online]. Available: \url{https://gdpr.eu/cookies/}
\BIBentrySTDinterwordspacing

\bibitem{Iqbal2021}
U.~Iqbal, S.~Englehardt, and Z.~Shafiq, ``{Fingerprinting the Fingerprinters:
  Learning to Detect Browser Fingerprinting Behaviors},'' in \emph{2021 IEEE
  Symposium on Security and Privacy (SP)}.\hskip 1em plus 0.5em minus
  0.4em\relax IEEE, 05 2021, pp. 1143--1161.

\bibitem{Acer2014}
\BIBentryALTinterwordspacing
G.~Acar. (2014) {Browser Fingerprinting and the Online-Tracking Arms Race}.
  Accessed: 2024-09-27. [Online]. Available:
  \url{https://www.esat.kuleuven.be/cosic/news/the-web-never-forgets-persistent-tracking-mechanisms-in-the-wild/}
\BIBentrySTDinterwordspacing

\bibitem{ReqHead24}
\BIBentryALTinterwordspacing
``Request header,'' accessed: 2024-09-27. [Online]. Available:
  \url{https://developer.mozilla.org/en-US/docs/Glossary/Request_header}
\BIBentrySTDinterwordspacing

\bibitem{Http224}
\BIBentryALTinterwordspacing
``{HTTP/2 fingerprinting: A relatively-unknown method for web
  fingerprinting},'' accessed: 2024-09-27. [Online]. Available:
  \url{https://lwthiker.com/networks/2022/06/17/http2-fingerprinting.html}
\BIBentrySTDinterwordspacing

\bibitem{Laperdrix2016}
P.~Laperdrix, W.~Rudametkin, and B.~Baudry, ``{Beauty and the Beast: Diverting
  Modern Web Browsers to Build Unique Browser Fingerprints},'' in \emph{2016
  IEEE Symposium on Security and Privacy (SP)}, 2016, pp. 878--894.

\bibitem{UserAg24}
\BIBentryALTinterwordspacing
``{User-Agent},'' accessed: 2024-09-27. [Online]. Available:
  \url{https://developer.mozilla.org/en-US/docs/Web/HTTP/Headers/User-Agent}
\BIBentrySTDinterwordspacing

\bibitem{Wolford2024}
\BIBentryALTinterwordspacing
B.~Wolford. (2024) {What are the GDPR consent requirements?} Accessed:
  2024-09-27. [Online]. Available:
  \url{https://gdpr.eu/gdpr-consent-requirements/}
\BIBentrySTDinterwordspacing

\bibitem{Navi24}
\BIBentryALTinterwordspacing
``{Navigator: plugins property},'' accessed: 2024-09-27. [Online]. Available:
  \url{https://developer.mozilla.org/en-US/docs/Web/API/Navigator/plugins}
\BIBentrySTDinterwordspacing

\bibitem{Extdetect24}
\BIBentryALTinterwordspacing
``{Extension Detector},'' accessed: 2024-09-27. [Online]. Available:
  \url{https://github.com/z0ccc/extension-detector}
\BIBentrySTDinterwordspacing

\bibitem{FiBlocker24}
\BIBentryALTinterwordspacing
``{How ad blockers can be used for browser fingerprinting},'' accessed:
  2024-09-27. [Online]. Available:
  \url{https://fingerprint.com/blog/ad-blocker-fingerprinting/}
\BIBentrySTDinterwordspacing

\bibitem{Karami2020}
S.~Karami, P.~Ilia, K.~Solomos, and J.~Polakis, ``{Carnus: Exploring the
  Privacy Threats of Browser Extension Fingerprinting},'' in \emph{{27th Annual
  Network and Distributed System Security Symposium, {NDSS} 2020, San Diego,
  California, USA, February 23-26, 2020}}.\hskip 1em plus 0.5em minus
  0.4em\relax The Internet Society, 2020.

\bibitem{CanvasBlocker24}
\BIBentryALTinterwordspacing
``{CanvasBlocker},'' accessed: 2024-09-27. [Online]. Available:
  \url{https://github.com/kkapsner/CanvasBlocker}
\BIBentrySTDinterwordspacing

\bibitem{WebGL24}
\BIBentryALTinterwordspacing
``{WebGL: 2D and 3D graphics for the web},'' accessed: 2024-09-27. [Online].
  Available: \url{https://developer.mozilla.org/en-US/docs/Web/API/WebGL_API}
\BIBentrySTDinterwordspacing

\bibitem{Stephenson2023}
T.~Stephenson, ``{A Comparative Study on Analyses of Browser Fingerprinting},''
  Ph.D. dissertation, Wesleyan University, 2023.

\bibitem{WebKit24}
\BIBentryALTinterwordspacing
``{WebKit},'' accessed: 2024-09-27. [Online]. Available:
  \url{https://github.com/WebKit/WebKit/commit/ae710d34c23858295b385e3f95ad7f6edd29f9d7}
\BIBentrySTDinterwordspacing

\bibitem{Nakibly2015}
G.~Nakibly, G.~Shelef, and S.~Yudilevich, ``{Hardware Fingerprinting Using
  HTML5},'' \emph{arXiv preprint arXiv:1503.01408}, 03 2015.

\bibitem{Cao2017CrossBrowserFV}
Y.~Cao, S.~Li, and E.~Wijmans, ``{(Cross-)Browser Fingerprinting via OS and
  Hardware Level Features},'' in \emph{Network and Distributed System Security
  Symposium}, 2017.

\bibitem{Mantel2022}
\BIBentryALTinterwordspacing
M.~Mantel. (2022) {Browser-Fingerprinting: PCs, Smartphones \& Co. lassen sich
  über die GPU tracken}. Accessed: 2024-09-27. [Online]. Available:
  \url{https://www.heise.de/news/Browser-Fingerprinting-PCs-Smartphones-Co-lassen-sich-ueber-die-GPU-tracken-6345233.html}
\BIBentrySTDinterwordspacing

\bibitem{Laor2022DRAWNAPARTAD}
\BIBentryALTinterwordspacing
{Laor et al.}, ``{DRAWNAPART: A Device Identification Technique based on Remote
  GPU Fingerprinting},'' \emph{ArXiv}, vol. abs/2201.09956, 2022. [Online].
  Available: \url{https://api.semanticscholar.org/CorpusID:246276013}
\BIBentrySTDinterwordspacing

\bibitem{Queiroz2019AWB}
\BIBentryALTinterwordspacing
J.~S. Queiroz and E.~L. Feitosa, ``{A Web Browser Fingerprinting Method Based
  on the Web Audio API},'' \emph{Comput. J.}, vol.~62, pp. 1106--1120, 2019.
  [Online]. Available: \url{https://api.semanticscholar.org/CorpusID:86644316}
\BIBentrySTDinterwordspacing

\bibitem{Chalise2021}
S.~Chalise and P.~Vadrevu, ``{A Study of Feasibility and Diversity of Web Audio
  Fingerprints},'' \emph{arXiv preprint arXiv:2107.14201}, 2021.

\bibitem{Gomez2018}
\BIBentryALTinterwordspacing
A.~G\'{o}mez-Boix, P.~Laperdrix, and B.~Baudry, ``{Hiding in the Crowd: an
  Analysis of the Effectiveness of Browser Fingerprinting at Large Scale},'' in
  \emph{Proceedings of the 2018 World Wide Web Conference}, ser. WWW '18.\hskip
  1em plus 0.5em minus 0.4em\relax Republic and Canton of Geneva, CHE:
  International World Wide Web Conferences Steering Committee, 2018, p.
  309–318. [Online]. Available: \url{https://doi.org/10.1145/3178876.3186097}
\BIBentrySTDinterwordspacing

\bibitem{Englehardt2016}
\BIBentryALTinterwordspacing
S.~Englehardt and A.~Narayanan, ``{Online Tracking: A 1-million-site
  Measurement and Analysis},'' in \emph{Proceedings of the 2016 ACM SIGSAC
  Conference on Computer and Communications Security}, ser. CCS '16.\hskip 1em
  plus 0.5em minus 0.4em\relax New York, NY, USA: Association for Computing
  Machinery, 2016, p. 1388–1401. [Online]. Available:
  \url{https://doi.org/10.1145/2976749.2978313}
\BIBentrySTDinterwordspacing

\bibitem{FontAPI24}
\BIBentryALTinterwordspacing
``{Local Font Access API},'' accessed: 2024-09-27. [Online]. Available:
  \url{https://developer.mozilla.org/en-US/docs/Web/API/Local_Font_Access_API}
\BIBentrySTDinterwordspacing

\bibitem{Laperdrix2020}
\BIBentryALTinterwordspacing
P.~Laperdrix, N.~Bielova, B.~Baudry, and G.~Avoine, ``{Browser Fingerprinting:
  A Survey},'' \emph{ACM Trans. Web}, vol.~14, no.~2, apr 2020. [Online].
  Available: \url{https://doi.org/10.1145/3386040}
\BIBentrySTDinterwordspacing

\bibitem{Reiter2017}
\BIBentryALTinterwordspacing
A.~Reiter and A.~Marsalek, ``{WebRTC: your privacy is at risk},'' in
  \emph{Proceedings of the Symposium on Applied Computing}, ser. SAC '17.\hskip
  1em plus 0.5em minus 0.4em\relax New York, NY, USA: Association for Computing
  Machinery, 2017, p. 664–669. [Online]. Available:
  \url{https://doi.org/10.1145/3019612.3019844}
\BIBentrySTDinterwordspacing

\bibitem{DetectRTC24}
\BIBentryALTinterwordspacing
``{DetectRTC},'' accessed: 2024-09-27. [Online]. Available:
  \url{https://github.com/muaz-khan/DetectRTC}
\BIBentrySTDinterwordspacing

\bibitem{Bernardo2016}
\BIBentryALTinterwordspacing
V.~Bernardo and D.~Domingos, ``{Web-based Fingerprinting Techniques},'' in
  \emph{Proceedings of the 13th International Joint Conference on E-Business
  and Telecommunications}, ser. ICETE 2016.\hskip 1em plus 0.5em minus
  0.4em\relax Setubal, PRT: SCITEPRESS - Science and Technology Publications,
  Lda, 2016, p. 271–282. [Online]. Available:
  \url{https://doi.org/10.5220/0005965602710282}
\BIBentrySTDinterwordspacing

\bibitem{Bujlow2017}
T.~Bujlow, V.~Carela-Español, J.~Solé-Pareta, and P.~Barlet-Ros, ``{A Survey
  on Web Tracking: Mechanisms, Implications, and Defenses},'' \emph{Proceedings
  of the IEEE}, vol. 105, no.~8, pp. 1476--1510, 2017.

\bibitem{Olejnik2012}
\BIBentryALTinterwordspacing
L.~Olejnik, C.~Castelluccia, and A.~Janc, ``{Why Johnny Can't Browse in Peace:
  On the Uniqueness of Web Browsing History Patterns},'' \emph{12th Privacy
  Enhancing Technologies Symposium (PETS 2012)}, 07 2012. [Online]. Available:
  \url{https://petsymposium.org/2012/papers/hotpets12-4-johnny.pdf}
\BIBentrySTDinterwordspacing

\bibitem{Takei2015}
N.~Takei, T.~Saito, K.~Takasu, and T.~Yamada, ``{Web Browser Fingerprinting
  Using Only Cascading Style Sheets},'' in \emph{2015 10th International
  Conference on Broadband and Wireless Computing, Communication and
  Applications (BWCCA)}, 2015, pp. 57--63.

\bibitem{CSSFingerp24}
\BIBentryALTinterwordspacing
``{CSS-Fingerprint},'' accessed: 2024-09-27. [Online]. Available:
  \url{https://github.com/OliverBrotchie/CSS-Fingerprint}
\BIBentrySTDinterwordspacing

\bibitem{AdvTorFiP24}
\BIBentryALTinterwordspacing
``{Advanced Tor Browser Fingerprinting},'' accessed: 2024-09-27. [Online].
  Available:
  \url{http://jcarlosnorte.com/security/2016/03/06/advanced-tor-browser-fingerprinting.html}
\BIBentrySTDinterwordspacing

\bibitem{GetClient24}
\BIBentryALTinterwordspacing
``{Investigate impact of fingerprinting via getClientRects()},'' accessed:
  2024-09-27. [Online]. Available:
  \url{https://gitlab.torproject.org/tpo/applications/tor-browser/-/issues/18500}
\BIBentrySTDinterwordspacing

\bibitem{Wang2021}
H.~Wang, H.~Sayadi, A.~Sasan, P.~D. Sai~Manoj, S.~Rafatirad, and H.~Homayoun,
  ``{Machine Learning-Assisted Website Fingerprinting Attacks with Side-Channel
  Information: A Comprehensive Analysis and Characterization},'' in \emph{2021
  22nd International Symposium on Quality Electronic Design (ISQED)}, 2021, pp.
  79--84.

\bibitem{WebAssembly24}
\BIBentryALTinterwordspacing
``{WebAssembly},'' accessed: 2024-09-27. [Online]. Available:
  \url{https://developer.mozilla.org/en-US/docs/WebAssembly}
\BIBentrySTDinterwordspacing

\bibitem{PrecTiming24}
\BIBentryALTinterwordspacing
``{High precision timing},'' accessed: 2024-09-27. [Online]. Available:
  \url{https://developer.mozilla.org/en-US/docs/Web/API/Performance_API/High_precision_timing}
\BIBentrySTDinterwordspacing

\bibitem{DSGVOFP24}
\BIBentryALTinterwordspacing
``{Browser Fingerprinting und das TDDDG: Erlaubt oder nicht? [Browser
  Fingerprinting and the TDDDG: Allowed or not?]},'' accessed: 2024-09-27.
  [Online]. Available:
  \url{https://dr-dsgvo.de/browser-fingerprinting-und-das-ttdsg/}
\BIBentrySTDinterwordspacing

\bibitem{Pugliese2020}
G.~Pugliese, C.~Riess, F.~Gassmann, and Z.~Benenson, ``{Long-Term Observation
  on Browser Fingerprinting: Users' Trackability and Perspective},''
  \emph{Proceedings on Privacy Enhancing Technologies}, vol. 2020, pp.
  558--577, 05 2020.

\bibitem{Al-Fannah2020}
N.~Al-Fannah and C.~Mitchell, ``{Too little too late: can we control browser
  fingerprinting?}'' \emph{Journal of Intellectual Capital}, vol.
  ahead-of-print, 01 2020.

\bibitem{Kollnig2022}
\BIBentryALTinterwordspacing
K.~Kollnig, A.~Shuba, M.~Van~Kleek, R.~Binns, and N.~Shadbolt, ``{Goodbye
  Tracking? Impact of iOS App Tracking Transparency and Privacy Labels},'' in
  \emph{Proceedings of the 2022 ACM Conference on Fairness, Accountability, and
  Transparency}, ser. FAccT '22.\hskip 1em plus 0.5em minus 0.4em\relax New
  York, NY, USA: Association for Computing Machinery, 2022, p. 508–520.
  [Online]. Available: \url{https://doi.org/10.1145/3531146.3533116}
\BIBentrySTDinterwordspacing

\bibitem{Lin2023}
X.~Lin, F.~Araujo, T.~Taylor, J.~Jang, and J.~Polakis, ``{Fashion Faux Pas:
  Implicit Stylistic Fingerprints for Bypassing Browsers' Anti-Fingerprinting
  Defenses},'' in \emph{2023 IEEE Symposium on Security and Privacy (SP)},
  2023, pp. 987--1004.

\end{thebibliography}
%\bibliography{references.bib} 
%\printbibliography

%\begingroup
%\sloppy
%\printbibliography
%\endgroup 

\end{document}